\documentclass{article}
\usepackage[utf8]{inputenc}

\usepackage{amsmath,amssymb}

\usepackage{multirow}
\usepackage{placeins}

\usepackage{cite}
\usepackage{nameref,hyperref}
\makeatletter
\renewcommand{\@biblabel}[1]{\quad#1.}
\makeatother

\title{Power Structure in Chilean News Media}

\author{Jorge Bahamonde\textsuperscript{1},
Johan Bollen\textsuperscript{2},
Erick Elejalde\textsuperscript{3,*},\\
Leo Ferres\textsuperscript{4},
Barbara Poblete\textsuperscript{1}}
\date{October 2017}

\usepackage{graphicx}

\begin{document}

\maketitle

\noindent{1} Department of Computer Science, University of Chile, Santiago, Chile
\\
{2} School of Informatics and Computing, Indiana University, Bloomington, IN, USA
\\
{3} Computer Science Department, Faculty of Engineering, Universidad de Concepción, Concepción, Chile
\\
{4} Institute of Data Science, Faculty of Engineering, Universidad del Desarrollo \& Telefónica R\&D, Santiago, Chile
\\
\bigskip

\noindent* Corresponding author: E-mail: eelejalde@udec.cl (EE) 

\noindent Authors listed in alphabetical order. 

\noindent All authors contributed equally to this work.

\section*{Abstract}
Even democracies endowed with the most active free press struggle to maintain diversity of news coverage. Consolidation and market forces may cause only a few dominant players to control the news cycle. Editorial policies may be biased by corporate ownership relations, narrowing news coverage and focus. To an increasing degree this problem also applies to social media news distribution, since it is subject to the same socio-economic drivers. To study the effects of consolidation and ownership on news diversity, we model the diversity of Chilean coverage on the basis of ownership records and social media data. We create similarity networks of news outlets on the basis of their ownership and the topics they cover. We then examine the relationships between the topology of ownership networks and content similarity to characterize how ownership affects news coverage.  A network analysis reveals that Chilean media is highly concentrated both in terms of ownership as well as in terms of topics covered. Our method can be used to determine which groups of outlets and ownership exert the greatest influence on news coverage.

\section*{Introduction}
Chomsky once commented on the role of a free and diverse press: ``The smart way to keep people passive and obedient is to strictly limit the spectrum of acceptable opinion, but allow very lively debate within that spectrum.'' \cite{chomsky1998common} This is in fact
the position that many advanced democracies find themselves in as the diversity of news coverage seems to shrink, whereas news coverage itself seems to continuously expand in a non-stop news cycle. For example, the US has gone from 50 companies in 1983 to only 6 companies that control 90\% of media outlets in 2000, with further consolidation possible in the future\cite{bagdikian2004new}. Lack of diversity of viewpoints, topics, and representation of communities has been attributed to this relentless process of consolidation \cite{Winseck2008}.

The emerging lack of diversity and coverage in news reporting is frequently attributed to two specific factors.
First, as news media outlets attempt to cater to their audiences or community, they narrow their coverage to community-specific material, and as a consequence further limit their audience awareness in a homophilic cycle of mutual preferential attachment. This effect has recently been studies in terms of so-called online ``filter bubbles'' \cite{pariser2011filter}, in which users can choose to subscribe to outlets and news that confirm their pre-existing preferences and view-points, not only narrowing their own media exposure, but also encouraging outlets to increasingly specialize to smaller and more defined communities \cite{bakshy2015exposure}.

Second, the market-driven consolidation of the news media industry may lead to concentration of ownership. According to Chomsky's Propaganda model\cite{chomsky1988consent}, this concentration of ownership may have direct and indirect effects on editorial policies. For example, the editorial board of a newspaper owned by a group that also invests in agriculture may perceive a pressure to report more favorably about agricultural initiatives. Unlike other factors such as the (frequently explicitly publicized) political and historical mission of the outlet and its readership, ownership bias may thus exert a more insidious effect on editorial policies that is difficult to operationalize and quantify. Nevertheless, it may have a significant effect on the degree to which news consumers perceive the world and their ability to gather objective and effective information.

With the emergence of online social media platforms, most news outlets, from the smallest to the largest, have established an online presence that they use for real-time distribution of news content\cite{messner2011shovel}. These online environments provide an opportunity to test hypotheses with respect to the drivers affecting news diversity, such as consolidation, coverage, ownership, and network homophily. 

Twitter \cite{twitter} is a prime example of a social media platform geared towards the real-time distribution of news. Twitter enables its users to post short messages, called "tweets", that can be up to a maximum of 140 characters in length. Users can choose to subscribe to the tweets posted by other users. Once they ``follow'' a given user they will receive that user's Tweets in their own feed. Tweets can originate from individual users, but also from news outlets and other organizations. In fact, the large majority of news outlets post their most recent headlines and a brief summary of their news on their Twitter accounts in real-time. They are followed by large numbers of Twitter users who have made Twitter their primary news source.

The Twitter platform is particularly interesting as foundation for the study of news diversity and coverage since it is designed to constitute a large-scale social network. The ensemble of users-following-users establishes a social network where tweets travel along the edges of the network. This renders the social media platform an ideal laboratory to apply the toolkit of network science to the investigation of news diversity and coverage from a top-down (user to user to news outlet ) as well as a bottom-up perspective (news outlets to their followers).

In our current study we research the influence that ownership relations have on news media content and coverage by quantifying the strength of the relation between news media ownership and news media content diversity in Twitter. We analyze the user accounts of news media outlets to study how their content evolves and overlaps, and whether or not these observations are linked to their known ownership structure. 

We focus on Chilean news outlets since they have established a significant social media presence with a high number of Chilean users per 1000 individuals \cite{10.1371/journal.pone.0061981}. In addition, Chilean news has a clearly defined national audience which is geographically and culturally well-demarcated, thereby providing a distinct sample from previous media studies that were focused on English-speaking countries. The Chilean media landscape is furthermore well documented due to the availability of detailed, publicly available data with respect to its ownership structure, compiled by \emph{Poderopedia}, a journalist NGO that aims to understand power relationships between people, companies, and organizations.

We use the latter information to trace the existing ownership structure of Chilean media outlets which we then compare to the structural properties of their Twitter coverage and content, in particular with respect to the similarity of the content they publish in social media. To this end, we define a series of different content similarity metrics and evaluate for each one its relation to ownership.

Prior work has focused on studying story selection similarity within different news outlets \cite{10.1371/journal.pone.0014243}.  However, we extend this research by searching for indications of deeper interconnections in the \emph{mediasphere} at the intra-country level. An {\em et al.} \cite{citeulike:9609587} modeled the outline of digital media on Twitter, analyzing media similarity based on the degree of overlap between their respective follower communities.  They reported a strong tendency for members of the communities to read news from multiple sources, mostly on similar topics.  Park {\em et al.} \cite{Park:2012:CFM:2209310.2209311} proposed a system to identify and track events, in order to present different points of view of the same affair to readers to counteract opinion bias in news.  Saez-Trumpe {\em et al.} \cite{Saez-Trumper:2013:SMN:2505515.2505623} define a methodology to identify ``selection'' or ``gatekeeping''-bias which consist of editorial decisions to publish certain stories and not others.  They study these biases with respect to the prominence of the stories and the geographical location of the outlet.  Since their work uses a data set of media from different countries, they find that geography might influence the selection of the stories. 

Our work complements prior research by searching for potential causal pathways to explain the homophilic relations between groups of news outlets. This might help to identify and characterize potential filter bubbles, possibly informing novel recommendation system aimed at increasing the diversity of news consumption

\section*{Materials and methods}
\label{sec:mat_met}

Our goal is to analyze whether ownership and content are correlated in the domain of digital media news outlets.  We approach this problem by studying the similarity networks and clusters that emerge from the content published on Twitter by news outlets in Chile. We contrast groups of similar news accounts with their ownership in the real-world to see if they are related according to different similarity metrics.

In particular, we study the similarity between pairs of news accounts from several perspectives: {\em vocabulary}, {\em keyword-based topics}, and {\em minhash-based topics}. We aim to determine if there exist consistent similarity-based communities among news media outlets and if this same consistency arises in relation to ownership.

In order to achieve this, we perform independent static analyses of news media outlets for two years, 2015 and 2016. For each year, we study the communities of news outlets that are produced by using community detection over similarity graphs built for each similarity metric. In addition, we identify clusters of similar outlets with the purpose of checking consistency of the resulting similarity groups. Below, we detail our similarity metrics, and community and clustering algorithms. Our data analysis started from the following:

\begin{itemize}
    \item{\bf Chilean News Twitter ({\em ds15}):}  all tweets published by $84$ prominent Chilean news media outlets from October 30th, 2014 through May 20th, 2015 (including retweets).  This data set contains $714,973$ tweets and was created by Maldonado {\em
et al.} \cite{jaz} for their study that characterized Chilean news events.
    \item{\bf Chilean News Twitter ({\em ds16}):} A manually curated and exhaustive list of news outlets in Chile for year 2016. This list derived from the Wikipedia page listing Chilean news media \cite{wiki.chilean.newspapers} and the independent journalistic website Poderopedia
\end{itemize}

We joined both sets and kept all of those that had an active Twitter account; then, we downloaded all of the tweets generated by those accounts from October 25, 2015 to January 25, 2016. Overall, the 
\textit{ds16} collection contained 365 news accounts and 756,864 tweets (also including retweets). Both data sets include  tweet metadata, such as location and user identifiers.

Standard text normalization and cleaning techniques were used to convert Tweet content in both data sets to lower case and remove stop-words, URLs, and punctuation. In addition, news outlets that posted less than one tweet per day on average were removed, leaving 79 news outlets for {\em ds15} and 341 in {\em ds16}.

As for ownership information, we manually mapped Poderopedia's influence database \cite{poderopedia}
to our lists of news media accounts on Twitter.  As for grouping news media outlets according to their owners, we simply consider two outlets to belong to the same group if and only if they're owned by the same entity.  There are at least two possible issues with this.  First, some news media outlets are owned by multiple entities: in this case, we selected the major partner. On
the other hand, there also exist ownership relationships {\em between} owners. In this case, we selected the entity that subsumes all others as the owner.  As a result we obtained the first complete database of newspaper ownership information in Chile (Url to be made available for the camera-ready version of this work).

Our datasets include news outlets that belong to the two biggest news media groups in Chile: the {\em El Mercurio} group and the {\em Copesa} media conglomerate, which form what has been called in the past a newspaper duopoly\cite{Castro2008}.  We also have representatives of a group of digital newspapers, the {\em Mi Voz} network. Other owners with smaller number of outlets are also included, as well as a group of {\em
unknown-to-us ownership}.  We note that we are interested not only 
in news outlets that share owners, but also those that behave as if they did.

\subsection*{Similarity metrics between news outlets} 

We are interested in finding how related news outlets are using their content. In other words, we study their vocabularies and how the stories they select may indicate a connection between their editorial policy. For topic-based similarity metrics, we decided to use keyword- and minhash-based similarity to validate the consistency of the results beyond the selected methodology.

\begin{enumerate}
\item{\bf Vocabulary-based similarity.} 
We model each
news media outlet as a single document composed of all of the tweets posted
by its news account during the time of our data collection. Each document is
converted to its vector-space representation
using a {\em
tf-idf} weighting scheme\cite{manning2008scoring}. Similarity is then
computed as the cosine similarity between two vectors\cite{singhal2001modern}.

\item{\bf Keyword-based topic similarity.}
This is a more elaborate notion of content-based similarity between news
media outlets, which is based on whether two sources effectively talk about the same
topics. 

We identify the topics that were discussed during each day in our dataset. Each topic is obtained by mining frequent term-sets from the tweets posted that day and then joining these sets by word co-occurrence (within the same day).

A daily vector representation is computed for each outlet based on the day's topics; daily similarities between pairs of outlets are obtained as the cosine similarities of pairs of these vectors. Finally, the overall topic similarity between two outlets is defined as their average daily similarity.

\item{\bf Minhash-based topic similarity.} 
This is an alternative similarity measure based on text, inspired by prior
work for identifying similar documents. We represent the text of each tweet posted
by a news outlet by its $k$-shingles, using word-based shingles and $k=3$ 
(this is based on prior findings that indicate that $k$ equal to 2 or 3 word shingles are appropriate for short documents \cite{broder1997resemblance,broder2000identifying,manku2007detecting}). We set $k=3$ to obtain a fine grain classifier identifying specific stories, rather than broader topics for which a smaller value of $k$ may have been chosen.

A $4$-min-wise hashing is applied
to each tweet representation, this results in compact summaries of documents that are effective for identifying similarity
\cite{broder1997resemblance,broder2000identifying,manber1994finding}.
Accordingly, we cluster tweets using their minhash similarity. We
refer to these clusters as {\em topics}, as this is a notion that 
has been used in past literature
to identify ``stories'' that relate to a common event or topic among news
outlets \cite{10.1371/journal.pone.0014243,Saez-Trumper:2013:SMN:2505515.2505623}.
Similarity between two news outlets is then defined by
the co-occurrence of two news outlets with respect to a same topic. 
In particular, the value of the similarity between outlet
$A$ and $B$ is the conditional probability $Pr(A|B)$ of the occurrence
of $A$ in a cluster given that $B$ occurs in that same cluster.  This
similarity measure is directional, expressing how likely it is that
a story tweeted by $B$ is also tweeted by $A$ \cite{citeulike:9609587}.
In order to define a symmetric similarity measure we further specify the
similarity between $A$ and $B$ as $sim(A,B)=max(Pr(A|B), Pr(B|A))$. 
\end{enumerate}

\subsection*{ Similar news outlet identification}

In order to identify similar news outlets, we create different {\em similarity
graphs} based on each of the aforementioned similarity metrics and each of the datasets. Formally, we define
a generic similarity graph $G=(V, S)$ for the set of news outlets
$V=\{v_1, v_2, \dots, v_n\}$ and similarity measure $S:(v_i,v_j)
\rightarrow {\rm I\!R^+}$, as a graph where each pair of outlets $v_i$ and
$v_j$ are connected by an edge of weight $S:(v_i,v_j)$. This yields a complete,
weighted, and undirected graph.

Using each similarity graph, we apply graph partitioning techniques to find
groups of similar news outlets. For all six similarity graphs we used a hierarchical, agglomerative community 
discovery algorithm \cite{newman2004}, and the normalized cut technique \cite{Shi:2000:NCI:351581.351611}. 
This methodology has
been proved to be successful in similar problems \cite{Lu:2015:BLS:2806416.2806573}.

\section*{Results and discussion}
\label{sec:results}

\begin{figure}[ht]
 \centering
        \caption{\bf {\em ds15} vs. {\em ds16} communities on {\em Topic (keyword-based)} similarity. 
        }
        \includegraphics[width=.95\linewidth]{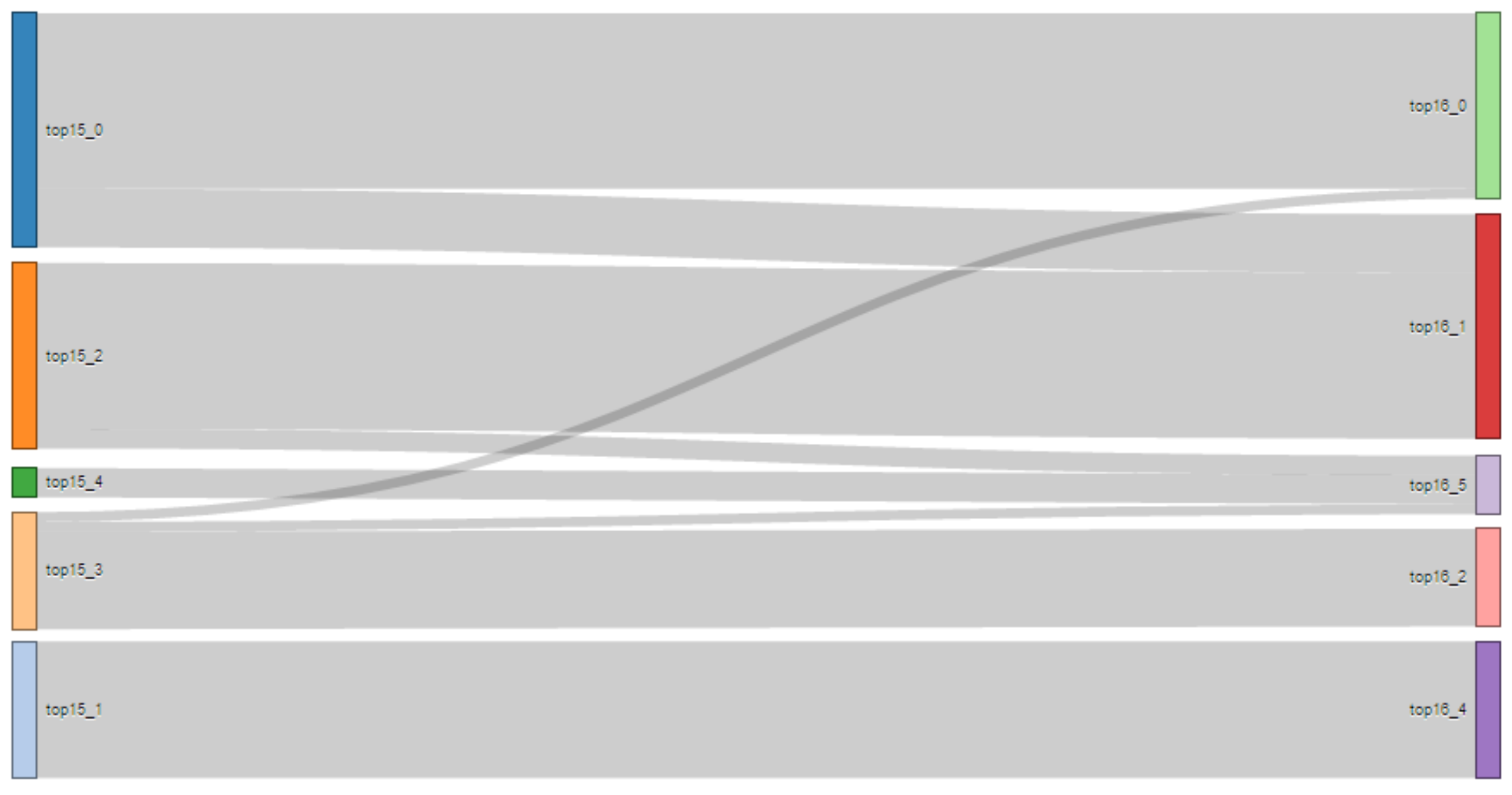}
        \label{fig:comm_topic_2015_2016}
\end{figure}

\begin{figure}[ht]
 \centering
        \caption{\bf Communities vs. Clusters {\em Topic (keyword-based)} similarity on {\em ds16}.}
        \includegraphics[width=.95\linewidth]{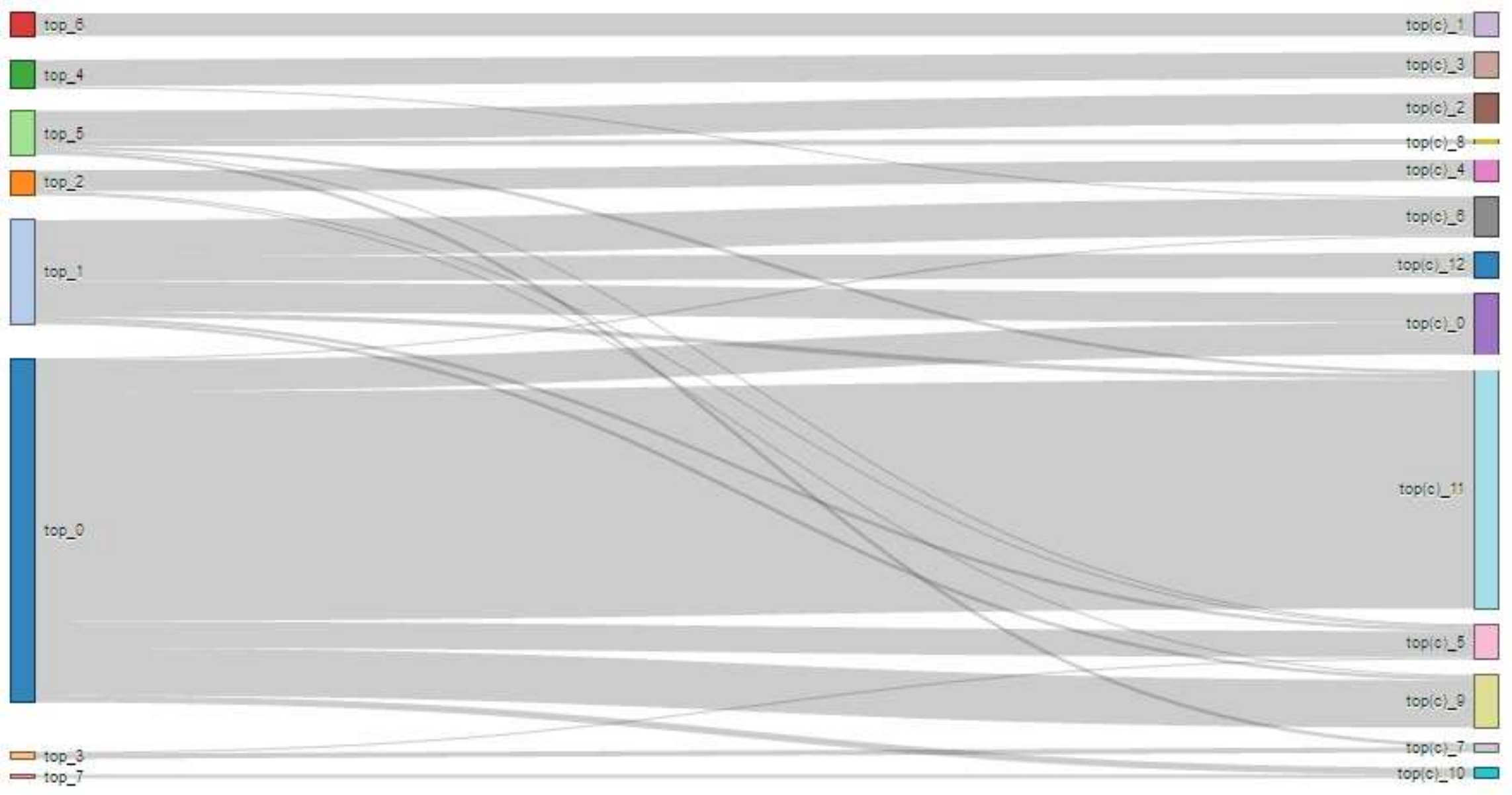}
        \label{fig:comm_clus_topic_2016}
\end{figure}

Figure \ref{fig:comm_topic_2015_2016} shows the communities (represented as boxes) of news outlets obtained from Topics collected in {\em ds15} (left column) and {\em ds16} (right column). The curves that connect both columns of boxes represent the number of outlets (or the proportion of outlets) shared by the communities at the two ends of the curve. For example, the first community in the top left (from {\em ds15}) has 18 (3/4) outlets in common with the first community in the top right (from {\em ds16}) and shares only 6 (1/4) outlets with the second. Also, the last community in left is identical to the last community in the right. Since results do not vary significantly between the two datasets, and to save space, we only report results for one of the datasets; the more recent one, {\em ds16}. In the same vein, we only report results for one of the graph partitioning algorithms (community detection) since the resulting communities in both cases are very similar. This congruence supports the notion that the communities found are significant, denoting a real structure in the data (see Fig \ref{fig:comm_clus_topic_2016}). In any case, the rest of the analysis can be found online at \url{https://github.com/eelejalde/Chilean_Media_Power_Structure.git} and in the Supporting Information section of this paper. This notwithstanding, there were some cases for which differences were found, and we do report these, making it clear where they come from.

Table \ref{table:summary_com} summarizes several metrics for community discovery over both data sets. The column {\em Outlets} is the initial number of outlets in the similarity graph (see Section \nameref{sec:mat_met}), while {\em Grouped} is the number of outlets that were included in one community. Column \textit{Comm.} is the number of communities found by the algorithm, while \textit{Mod.} and \textit{Cond.} show, respectively, the modularity and conductance of the sub-graphs formed by the \textit{Grouped} outlets within returned communities (see Section \nameref{sec:mat_met}).

\begin{table}[h!] 
\footnotesize
\caption{\bf Internal metrics for community structures derived from each explored similarity measure for the {\em ds15} and {\em ds16} datasets.}
\begin{tabular}{|l|r|r|r|r|r|r|r|r|r|r|} 
\hline
 \multirow{2}{*}{\bf Similarity} & \multicolumn{2}{c|}{\bf Outlets} & \multicolumn{2}{c|}{\bf Grouped} & \multicolumn{2}{c|}{\bf Comm.} & \multicolumn{2}{c|}{\bf Mod.} & \multicolumn{2}{c|}{\bf Cond.} \\
  \cline{2-11}
  & {\em ds15}  & {\em ds16} & {\em ds15} & {\em ds16} & {\em ds15} & {\em ds16} & {\em ds15} & {\em ds16} & {\em ds15} & {\em ds16}\\
  \hline
 Vocabulary     & 79 & 341 & 52 & 262 & 7 & 14 & 0.38 & 0.38 & 0.02 & 0.35\\
 \hline
 Topics 	    & 79 & 341 & 50 & 133 & 4 & 7  & 0.60 & 0.58 & 0.11 & 0.26\\
 \hline
 MinHash        & 75 & 365 & 50 & 355 & 6 & 11 & 0.40 & 0.74 & 0.01 & 0.04\\
\hline
\end{tabular}
\begin{flushleft} Grouped outlets (Grouped) correspond to those belonging to a discovered community. Modularity (Mod.) and conductance (Cond.) are calculated with respect to this subgraph.
\end{flushleft}
\label{table:summary_com}
\end{table}

The first thing to notice is that {\em Topic} similarity creates the lowest number of communities. Also, this similarity for the Dataset {\em ds15} includes almost all outlets, but for {\em ds16} it only grouped about a third of the total dataset. We think this is because this similarity is a coarse-grained classification that only captures the strongest signals. If we focus on {\em ds15}, the {\em topic} similarity has the highest modularity, which means it creates well defined communities. However, we have to take into account that this dataset only contains 84 outlets that comprise most of the largest, most famous newspapers of the country. When we look at the {\em ds16} dataset we find a more diverse set of outlets (in size and content). In {\em ds16}, {\em Topic} similarity shows similar performance if un-grouped outlets are excluded. In turn, {\em MinHash} seems to be more sensitive to weaker signals, creating a more fine-grained classification. We can see this in the high modularity achieved with {\em ds16} in spite of having included most outlets. On the other hand, {\em Vocabulary} similarity has the lowest performance in both datasets, which gives us the intuition that there are no particularly strong differences in vocabulary between the analyzed outlets.

\subsection*{Vocabulary similarity}

Communities obtained for the {\em Vocabulary} similarity over the {\em ds16} dataset include four big communities and ten smaller ones. Most of the smaller communities form around geographical information, as their relevant words include city names and places. In some cases, the outlet names also denote a geographical region, as many regional outlets have location-related names: ``El Mercurio de Valparaíso'', for example, is based in the city of Valparaíso. Thus, for some communities, we can infer regional-focused content by manually inspecting the names of the outlets grouped within it. We also find small communities pointing to a particular topic, such as aquaculture, and a particular type of media, such as radio stations. This is possibly caused by the fact that these two characteristics heavily influence the respective outlets' vocabularies: aquaculture is a very specific activity with its own terms, and radio stations tend to use a more informal language than other media outlets (see Tables \ref{table:vocab_nps2016} in the Supporting Information section for more detail).

There is also a pair of big communities that seems to comprise many different outlets: the first has many national-scope media, while the other has many that seem oriented toward regional locations.  The characterization of ownership for the same set of communities can be seen in Table \ref{table:vocab_ownership2016}. Notice that over a half of the community with ID 4 is owned by El Mercurio. Similarly, over half of the community with ID 6 is owned by Grupo Diarios en Red. On the other hand, the other two big communities (IDs 2 and 7) do not have any owner that heavily dominates the group. For the rest of the communities, even when they are very small ones, it is difficult to find homogeneity in the ownership of the groups. This results suggest a low influence of ownership of a news outlet over the vocabulary they use.

\begin{table}[h!]
\footnotesize
\caption{\bf Ownership properties for Vocabulary-based communities for the {\em ds16} dataset.}
\begin{tabular}{|r|r|p{4cm}|r|r|}
\hline
\textbf{ID} &   \textbf{Size} & \textbf{Main owner(s)} & \textbf{Owner(s)\% [\#]} & \textbf{Unk. owner \% [\#]}\\
\hline
0 & 79 & Copesa                    & 6.33 [5]  & 25.32 [20]\\
\hline
1 &  3 & Red de Diarios Comunales  & 66.67 [2] & 0.00 [0] \\
\hline
2 & 91 & Copesa                    & 14.29 [13] & 4.40 [4] \\
\hline
\multirow{4}{*}3 & \multirow{4}{*}{4} & Empresa Periodistica y Radiodifusora Las Nieves & 25.00 [1] & \multirow{4}{*}{0.00 [0]} \\
  \cline{3-4}
  &    & Sociedad Periodistica de Aysen & 25.00 [1]& \\
  \cline{3-4}
  &    & Sociedad Editora y Periodistica La Verdad  & 25.00 [1] & \\
  \cline{3-4}
  &    & Sociedad El Patagon Domingo & 25.00 [1] & \\
  \hline
4 & 43 & El Mercurio & 55.81 [24] & 11.63 [5] \\
\hline
\multirow{2}{*}5 &  \multirow{2}{*}{2} & Comunicaciones Mia & 50.00 [1] & \multirow{2}{*}{0.00 [0]} \\
\cline{3-4}
  &    & Grupo Prisa & 50.00 [1]& \\
  \hline
\multirow{2}{*}6 & \multirow{2}{*}{23} & Grupo Diarios en Red &  56.52 [13] & \multirow{2}{*}{17.39 [4]} \\
\cline{3-4}
  &    & El Mercurio & 17.39 [4]& \\
  \hline
\multirow{2}{*}7 & \multirow{2}{*}{78} & Asesorias e Inversiones Comunidades Ciudadanas & 14.10 [11]& \multirow{2}{*}{29.49 [23]} \\
\cline{3-4}
  &    & El Mercurio & 11.54 [9]& \\
  \hline
\multirow{2}{*}8 &  \multirow{2}{*}{2} & Mono Manco & 50.00 [1]& \multirow{2}{*}{0.00 [0]} \\
\cline{3-4}
  &    & Camilo Montalban Araneda & 50.00 [1]& \\
  \hline
\multirow{4}{*}9 &  \multirow{4}{*}{6} & Sociedad Periodistica e Impresora el Labrador & 16.67 [1]& \multirow{4}{*}{33.33 [2]} \\
\cline{3-4}
  &    & Portal de Melipilla & 16.67 [1]& \\
  \cline{3-4}
  &    & Editora el Centro Empresa Periodistica & 16.67 [1]& \\
  \cline{3-4}
  &    & Antonio Puga & 16.67 [1]& \\
  \hline
10 & 2 & - & - & 100.00 [2]\\
\hline
\multirow{2}{*}{11} & \multirow{2}{*}{2} & Red de Diarios Comunales & 50.00 [1]& \multirow{2}{*}{0.00 [0]} \\
\cline{3-4}
  &    & Sociedad Periodistica Banic y Lancelloti & 50.00 [1]& \\
  \hline
12 & 2 & Tu Ciudad Virtual & 50.00 [1]& 50.00 [1]\\
\hline
\multirow{2}{*}{13} & \multirow{2}{*}{2} & Editec & 50.00 [1]& \multirow{2}{*}{0.00 [0]} \\
\cline{3-4}
  &    & Sociedad Medios Comunicaciones & 50.00 [1]&  \\
  \hline
14 & 2 & Grupo Prisa & 100.00 [2]& 0.00 [0]\\
\hline
\end{tabular}
\begin{flushleft} The community with an ID of 0 corresponds to un-grouped media
outlets. Entities owning over 10\% of the outlets in a community are listed next
to it.
\end{flushleft}
\label{table:vocab_ownership2016} 
\end{table}

\subsection*{Topic similarity: Keyword based}

The computed community structure has a big community containing many national and 
local-scope media outlets, whose most relevant keywords are mainly centered around political and
sports figures and current issues. A small community is particularly oriented towards one particular region, Valpara\'iso; analogously, another community seems to be focused on another region, Aconcagua Province. The remaining communities seem to lack a unifying theme, displaying topics related to general
advice, geographical entities and buzzwords that aim to capture audience
interest. 

When ownership comes into play, these remaining communities acquire meaning (see Table \ref{table:topic_ownership2016}). The communities with IDs 2, 4, 5 and 6
have each an entity owning over 80\% of them. In contrast to the results seen before for the {\em Vocabulary} similarity, this indicates that ownership might have an influence on
topics discussed.

\begin{table}[ht]
\footnotesize
\caption{\bf  Ownership properties for keyword-based communities for the {\em ds16} dataset.}
\begin{tabular}{|r|r|p{5cm}|r|r|} 
\hline
\textbf{ID} &   \textbf{Size} & \textbf{Main owner(s)} & \textbf{Owner(s)\% [\#]} & \textbf{Unk. owner \% [\#]}\\
\hline
0 &    208 & Grupo Copesa & 5.29 [11]& 25.96 [54] \\
\hline
1 &     59 & Copesa & 11.86 [7]& 6.78 [4]\\
\hline
2 &     14 & El Mercurio & 85.71 [12]& 14.29 [2]\\
\hline
\multirow{4}{*}3 & \multirow{4}{*}{4} & Medios de Consorcio Periodistico El Epicentro & 25.00 [1]& \multirow{4}{*}{0.00 [0]}\\
\cline{3-4}
  &        & Corporacion de Television de la Pontificia Universidad Catolica de Valparaiso & 25.00 [1] & \\
\cline{3-4}
  &        & Comunicaciones Pacifico  & 25.00 [1] & \\
\cline{3-4}
  &        & Radio Festival & 25.00 [1] & \\
\hline
4 &     16 & Asesorias e Inversiones Comunidades Ciudadanas & 93.75 [15]& 0.00 [0]\\
\hline
5 &     25 & El Mercurio & 96.00 [24]& 0.00 [0]\\
\hline
6 &     13 & Grupo Diarios en Red & 100.00 [13]& 0.00 [0]\\
\hline
7 &      2 & Patricio Gallardo Montenegro & 50.00 [1]& 50.00 [1]\\
\hline
\end{tabular}
\begin{flushleft} The community with an ID of 0 corresponds to un-grouped media
outlets. Entities owning over 10\% of the outlets in a community are listed next
to it.
\end{flushleft}
\label{table:topic_ownership2016} 
\end{table}

Outlets owned by groups like {\em El Mercurio} or {\em Diarios en Red} have recognizable clusters using the agglomerative community detection algorithm. The {\em Copesa} group (the closest competitor of {\em El Mercurio}) does not have its own, clearly defined, community. This may be due to the kind of outlets it owns (a lot of them are magazines specialized in different topics). This makes it harder to identify any owner influence on their editorial strategies, which seems to be a limitation of the current methodology.

\subsection*{Topic similarity: Minhash-based}

Using the \textit{minhash} technique over the tweets in {\em ds16}, we identified 100,774 topics that contain 438,353 tweets. In the case of
{\em ds15}, we identify 83,582 topics containing a total of 254,650
tweets. We looked for topics that had tweets from multiple news outlets.  In {\em ds16}, out of all the topics, 31,423
contained tweets from more than one news outlet (31.2\%) and in {\em ds15}
$17,211$ (20.6\%).  Using these topics, we used the
co-occurrences for each pair of news outlets to calculate their similarity (see Section \nameref{sec:mat_met}). We found that all news outlets co-occur at least once with some other news source, for both {\em ds16} and {\em ds15}.

\begin{figure}[ht]
 \centering
        \caption{\bf Similarity graph, using {\em Topic (MinHash-based)} similarity on {\em ds16}. Only representing edges with weight over (mean+2std). We assigned different colors to the biggest owners.}
        \includegraphics[width=.9\linewidth]{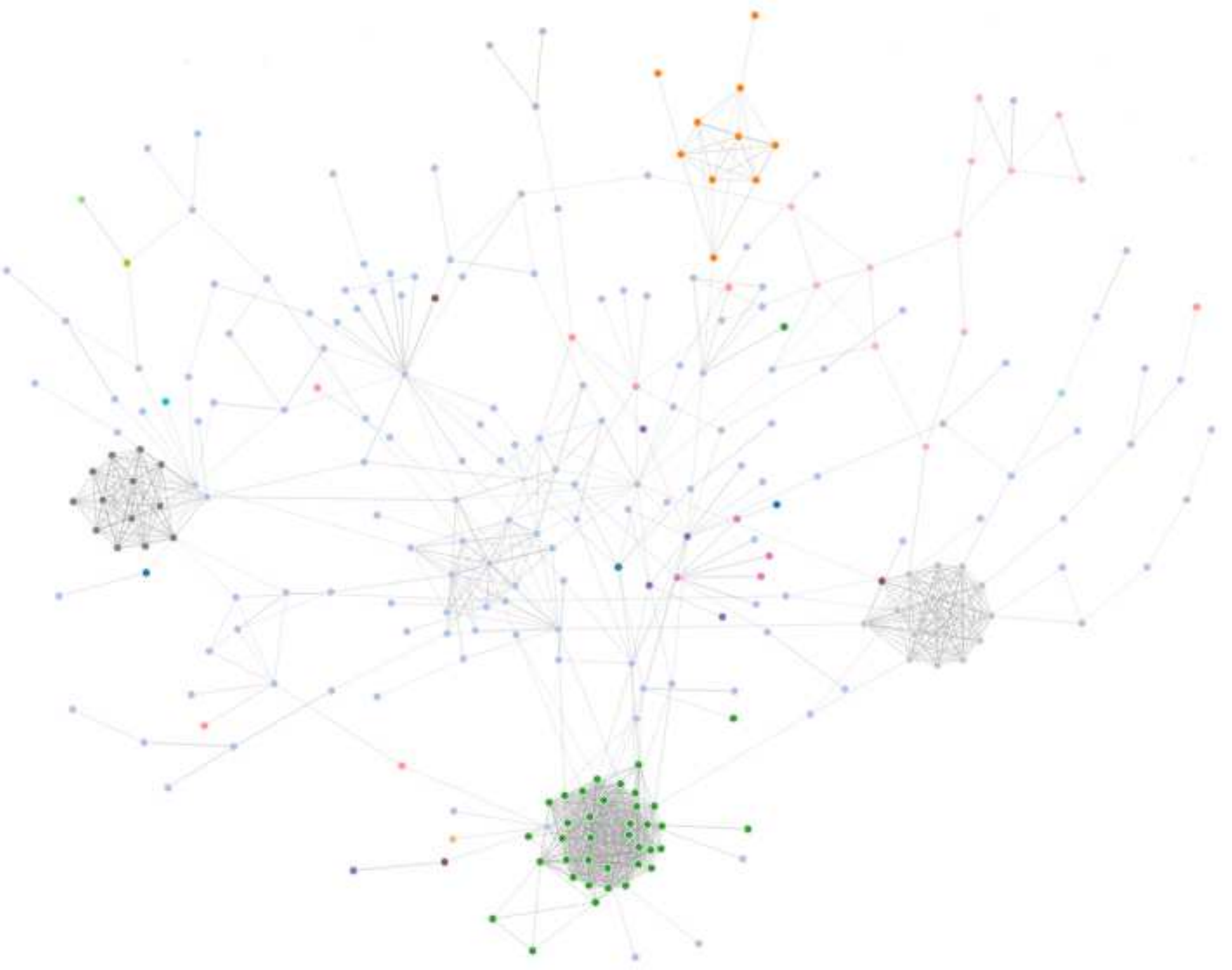}
        \label{fig:owners}
\end{figure}

Results for Minhash-based similarity have features like those seen in the
key\-word-based communities. These communities are easily identifiable even by visual inspection (see Fig \ref{fig:owners}). We observe two big communities with many different media outlets (with IDs 1 and 3), some small ones (with IDs of 4, 5, 10 and 11) which, as before, under manual inspection of the outlets name seem to have specific scopes (\textit{e.g.} the Linares Province, aquaculture, the Chiloe Province or the Maip\'u commune).

Ownership features help explain the remaining communities as shown in
Table \ref{table:minhash_ownership2016}. Communities 2, 6, 7 and 9 have entities that own a big part of them. The small community with an ID of 11 does not have a clear meaning or unifying theme. Notice that in this case the community with ID of 3 has similar characteristics to the communities with ID 0 for the other similarity measures (the un-grouped outlets). This community has the biggest number of outlets and the main owner has less than 10\% of them.

\begin{table}[ht]
\footnotesize
\caption{\bf Ownership properties for Minhash-based communities for the {\em ds16} dataset.}
\begin{tabular}{|r|r|p{5cm}|r|r|}
\hline
\textbf{ID} &   \textbf{Size} & \textbf{Main owner(s)} & \textbf{Owner(s)\% [\#]} & \textbf{Unk. owner \% [\#]}\\
\hline
0 &  10 & El Mercurio & 20.00 [2]& 30.00 [3]\\
\hline
1 & 109 & Red de Diarios Comunales & 12.84 [14]& 21.10 [23]\\
\hline
2 &  43 & El Mercurio & 83.72 [36]& 11.63 [5]\\
\hline
3 & 130 & Copesa & 4.62 [6]& 30.00 [39]\\
\hline
\multirow{2}{*}4 &  \multirow{2}{*}{2} & Radio Ancoa de Linares & 50.00 [1] & \multirow{2}{*}{0.00 [0]} \\
\cline{3-4}
  &    & Comunicaciones del Sur & 50.00 [1]&  \\
\hline
\multirow{2}{*}5 & \multirow{2}{*}{2} & Editec & 50.00 [1]& \multirow{2}{*}{0.00 [0]}\\
\cline{3-4}
  &	 	& Sociedad Medios Comunicaciones  & 50.00 [1]&  \\
\hline
6 &   9 & Betazeta Networks & 100.0 [9]& 0.00 [0]\\
\hline
7 &  36 & Asesorias e Inversiones Comunidades Ciudadanas & 41.67 [15]& 30.56 [10]\\
\hline
\multirow{4}{*}8 & \multirow{4}{*}{6} & El Mercurio & 16.67 [1] & \multirow{4}{*}{33.33 [2]} \\
\cline{3-4}
  & 	& Copesa  & 16.67 [1] & \\
\cline{3-4}
  &     & Troya Comunicaciones   & 16.67 [1] & \\
\cline{3-4}
  &     & Servicios de Radio Difusion Pedro Felidor Roa Barrientos   & 16.67 [1]& \\
\hline
9 &  14 & Grupo Diarios en Red & 92.86 [13]& 0.00 [0]\\
\hline
\multirow{2}{*}{10} &  \multirow{2}{*}{2} & Mono Manco & 50.00 [1]& \multirow{2}{*}{0.00 [0]}\\
\cline{3-4}
  &     & Camilo Montalban Araneda & 50.00 [1]& \\
\hline
11 &  2 & Sociedad Radiodifusora Primordial FM & 50.00 [1]& 50.00 [1]\\
\hline
\end{tabular}
\begin{flushleft} The community with an ID of 0 corresponds to un-grouped media
outlets. Entities owning over 10\% of the outlets in a community are listed next
to it.
\end{flushleft}
\label{table:minhash_ownership2016} 
\end{table}

Even though the communities in \textit{Minhash-based} similarity graph are partially explained by ownership, the correlation is not as strong as in the clustering obtained from the normalized cut algorithm. As mentioned above, this is one case where we think the differences between the two algorithms are worth mentioning: for this particular similarity graph the normalized cut clustering algorithm actually improves results (see Table \ref{table:cminhash_ownership2016}).

\begin{table}[h!]
\footnotesize
\caption{\bf Ownership properties for Minhash-based clustering for the {\em ds16} dataset.}
\begin{tabular}{|r|r|p{5cm}|r|r|}
\hline
\textbf{ID} &   \textbf{Size} & \textbf{Main owner(s)} & \textbf{Owner(s)\% [\#]} & \textbf{Unk. owner \% [\#]}\\
\hline
0 &  245 & Copesa & 5.71 [14]& 25.71 [63]\\
\hline
1 & 14 & - & -  & 50.00 [7]\\
\hline
2 &  19 & El Mercurio & 100.0 [19]& 0.00 [0]\\
\hline
3 & 16 & Asesorias e Inversiones Comunidades Ciudadanas & 93.75 [15]& 0.00 [0]\\
\hline
4 & 14 & El Mercurio & 100.0 [14] & 0.00 [0]\\
\hline
5 & 13 & Grupo Diarios en Red & 100.0 [13]& 0.00 [0]\\
\hline
6 & 9 & Grupo Prisa & 100.0 [9]& 0.00 [0]\\
\hline
7 &  4 & Editorial Televisa Chile & 100.0 [4]& 0.00 [0]\\
\hline
8 & 1 & Betazeta Networks & 100.0 [1]& 0.00 [0]\\
\hline
9 &  14 & Red de Diarios Comunales & 85.71 [12]& 0.00 [0]\\
\hline
\multirow{3}{*}{10} &  \multirow{3}{*}{3} & Estado de Chile & 33.33 [1]& \multirow{3}{*}{0.00 [0]} \\
\cline{3-4}
  &     & ITV Patagonia & 33.33 [1]& \\
\cline{3-4}
  &     & Corporacion de Television de la Pontificia Universidad Catolica de Valparaiso & 33.33 [1]& \\
\hline
\end{tabular}
\begin{flushleft} Entities owning over 10\% of the outlets in a cluster are listed next
to it.
\end{flushleft}
\label{table:cminhash_ownership2016} 

\end{table}

If we assume that the biggest cluster (with ID 0) is the one containing the outlets that do not fit in any other group (equivalent to the un-grouped outlets in the community detection), then we get clusters that are very similar to the communities we obtained for {\em topic (keyword-based)} similarity.

On one hand, the clusters leave out a bigger number of outlets than the community structure. This reduces the number of clustered outlets to an amount similar to that found with {\em topic (keyword-based)} similarity. On the other hand, it finds a classification with a better owner separation. As we can see in Table \ref{table:cminhash_ownership2016}, there are two relatively small clusters (with ID 8 and 10). Beside those two, all other clusters are heavily, if not entirely, dominated by one owner.

\subsection*{Clustering metrics}

Based on these results, we hypothesize that ownership relationships are similar to the ones based on
content. Given that we have the actual owners of most news outlets in our data
sets, we used this as a ground truth to evaluate the performance of our methodology. To this end, we computed different
clustering metrics using ownership information as class labels. 

We used the Adjusted Rand Index ({\bf ARI}) to quantify the degree of
correspondence between the set of communities found by our methodology and the sets
of clusters defined by the actual owners of the news outlets. {\bf
ARI} scores are normalized against chance, so scores close to $0.0$ indicate random
label assignments, $1.0$ indicates a perfect match, and negative scores indicate
a correspondence lower than what is expected for random assignments. Similarly,
the Adjusted Mutual Information ({\bf AMI}) index gives a sense of how much
information we can obtain about one distribution given the other one. {\bf
AMI} scores are also adjusted with respect to the expected value (subtracting the expected
value from the Mutual Information score). Again, scores close to $0.0$ indicate
random assignments and a $1.0$ score indicates two identical assignments. The
Normalized variation of the Mutual Information Index ({\bf NMI}) also gives a
greater score as the communities are closer to a perfect recreation of ownership
classes. Moreover, {\bf NMI} does not penalize if the classes are further
subdivided into smaller clusters. The results of the application of these
indices (given in Table \ref{table:clust_perf_eval}) suggest non-random
clusters. Homogeneity ({\bf Hom}) is maximized when each cluster contains
members of a single class, while completeness ({\bf Com}) measures the desirable
objective of assigning all members of a class to a single cluster.

Prior to calculations, we removed outlets without ownership information and
outlets with owners that only have a single outlet, since they do not add any relevant information. Additionally, there are outlets that do not belong to any of the communities we
found. They could be discarded, but we might be deleting valuable information:
our algorithm indicates their content is different from the others'. For this reason, we preserve each of them as a
community of size 1. Though reasonable, this might distort some comparison
metrics, as the correspondence of single-outlet communities is perfect if
they're isolated in both the content-based and the owner-based community
structures. As both the number of considered outlets and the number of
communities is altered by these decisions, we specify them in Table \ref{table:clust_perf_eval} (columns \textit{Outlets} and \textit{Comm.} respectively).

\begin{table}[ht]
\footnotesize
\caption{\textbf{Comparison of community structures and ownership.}}
\begin{tabular}{|l|r|r|r|r|r|r|r|}
    \hline
     Similarity   & Outlets & Comm. & ARI & AMI & NMI & Hom & Com \\
    \hline
     Vocabulary        & 157 & 28 & 0.1834 & 0.3007 & 0.5362 & 0.4748 & 0.6056 \\
    Topic: Keywords    & 157 & 66 & 0.4246 & 0.4261 & 0.7313 & 0.8113 & 0.6592 \\
     Topic: Minhash    & 167 & 12 & 0.4301 & 0.4584 & 0.6593 & 0.5460 & \textbf{0.7961} \\
     \hline
Cluster Topic: Minhash & 169 & 90 & \textbf{0.5326} & \textbf{0.4652} & \textbf{0.8365} & \textbf{1.0000} & 0.6997 \\
    \hline
\end{tabular}
\begin{flushleft} Rows represent the different metrics used to calculate the similarity graphs. Columns represent the scores of the indices calculated using our ground-truth as reference ({\bf ARI}: Adjusted Rand Index, {\bf AMI}: Adjusted Mutual Information Based, {\bf NMI}: Normalized Mutual Information Based, {\bf Hom}: Homogeneity, {\bf Com}: Completeness)
\end{flushleft}
\label{table:clust_perf_eval}
\end{table}

\begin{figure}[ht]
\caption{{\bf Ownership vs. Vocabulary community structure.}}
\centering
\includegraphics[width=0.75\textwidth]{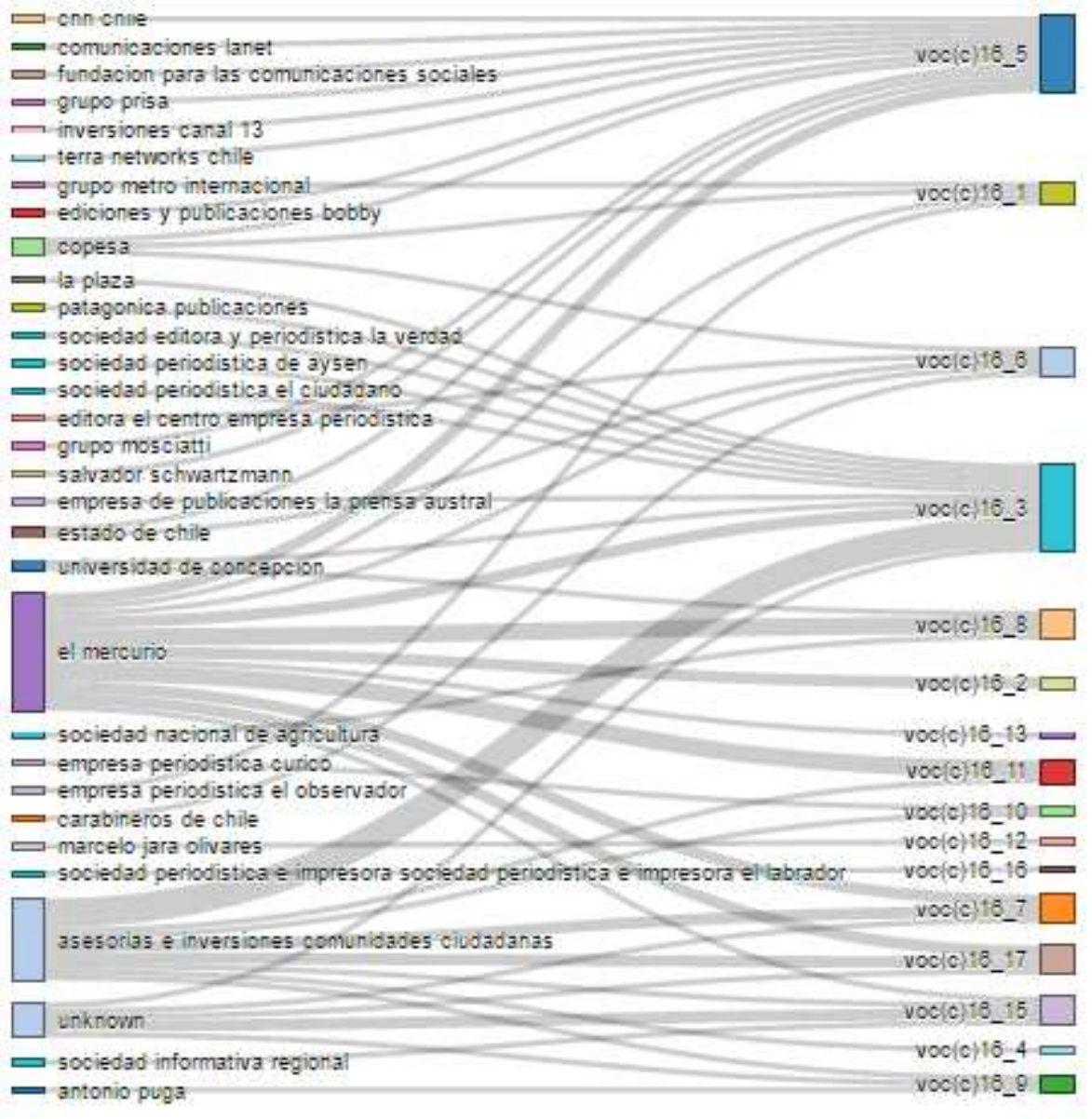}
\begin{flushleft}
Owners are displayed on the left, while communities are displayed on the right. The width of a flow connecting an owner and a community is proportional to the number of outlets in the community belonging to that owner.
\end{flushleft}
\label{comparison_vocabulary}
\end{figure}

\begin{figure}[ht]
\caption{{\bf Ownership vs. Topic (keyword-based) community structure.}}
\centering
\includegraphics[width=0.75\textwidth]{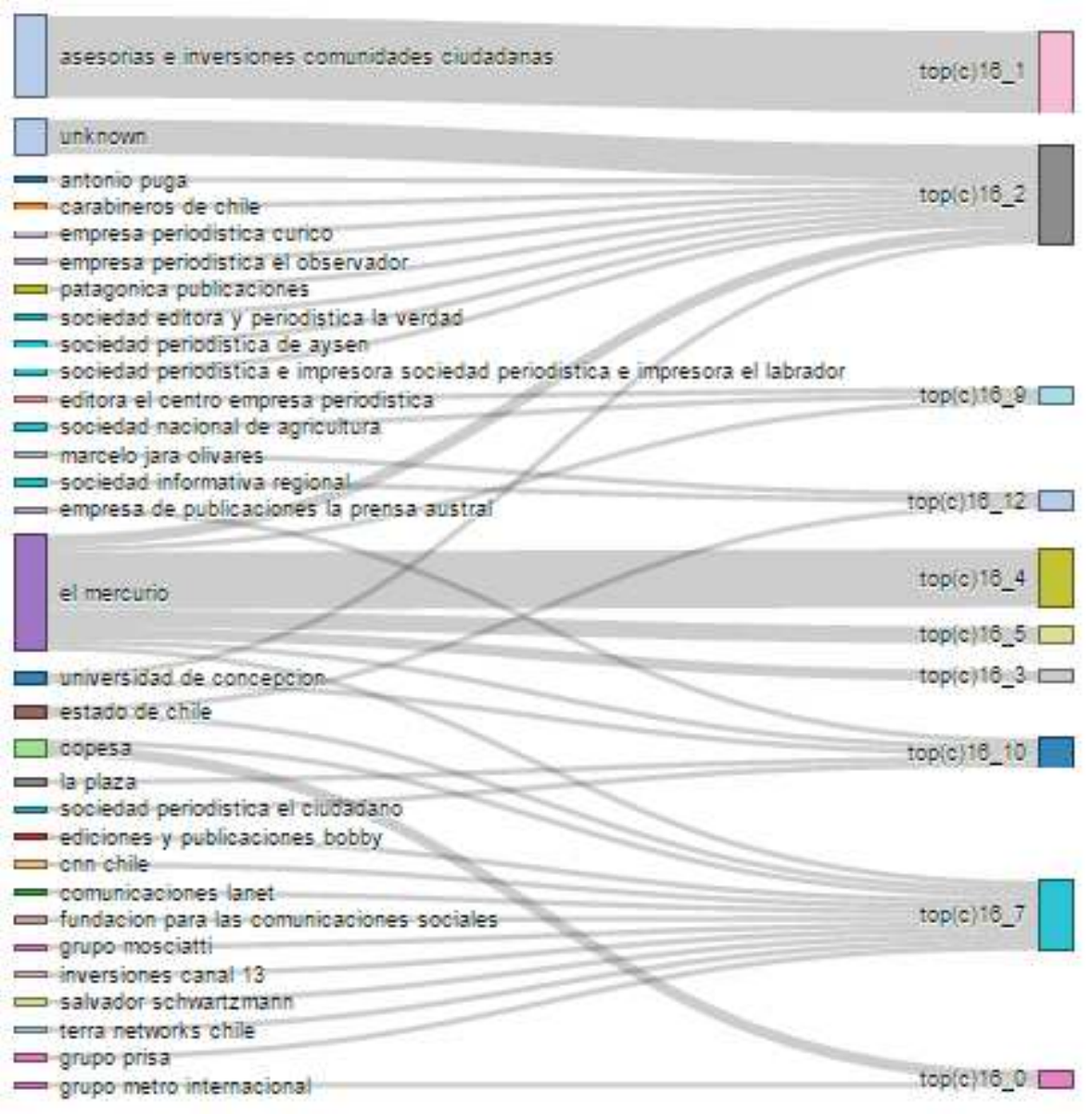}
\begin{flushleft}
Owners are displayed on the left, while communities are displayed on the right. The width of a flow connecting an owner and a community is proportional to the number of outlets in the community belonging to that owner.
\end{flushleft}
\label{comparison_topic}
\end{figure}

\begin{figure}[ht]
\caption{{\bf Ownership vs. Topic (minhash-based) community structure.}}
\centering
\includegraphics[width=0.75\textwidth]{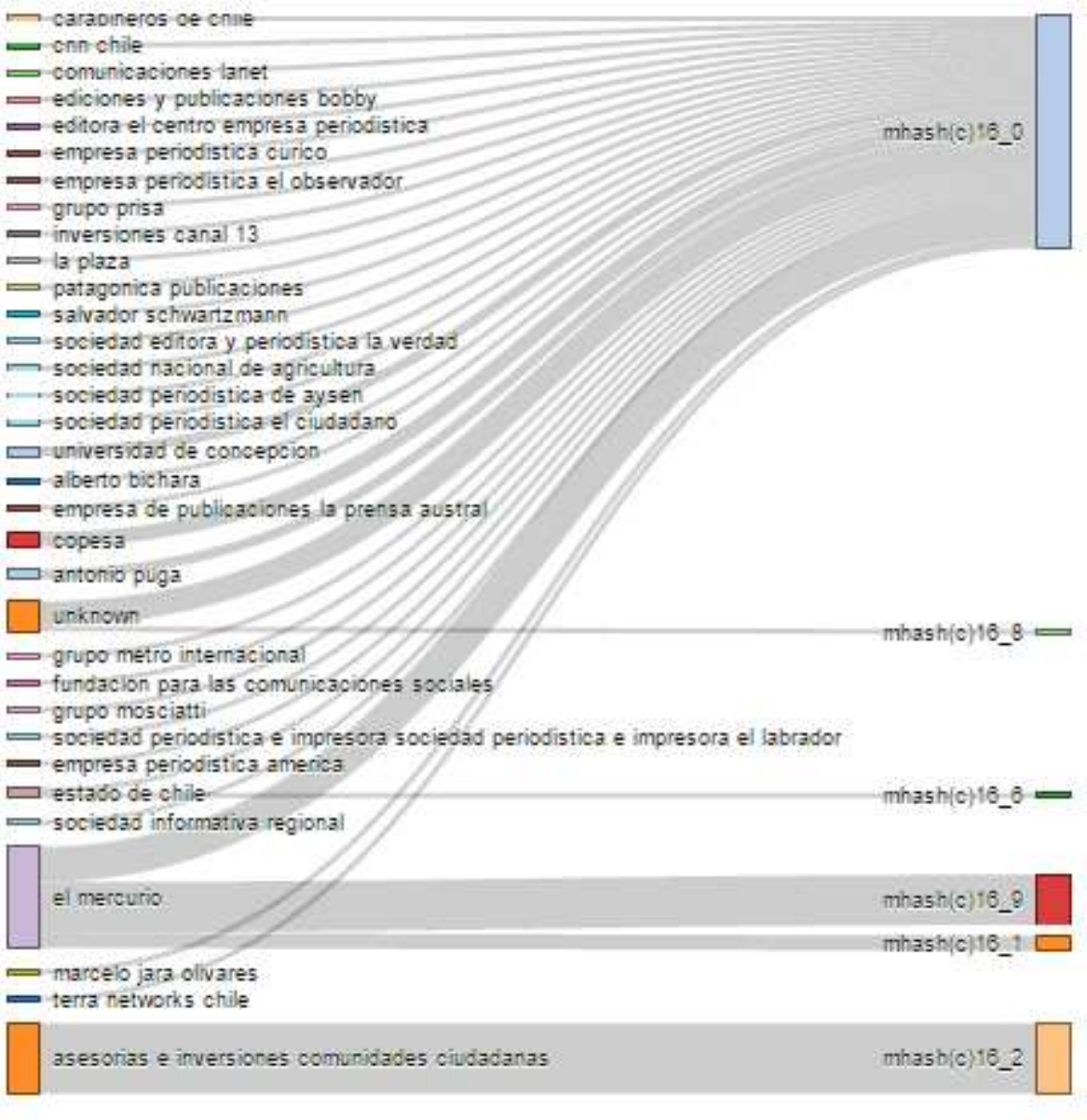}
\begin{flushleft}
Owners are displayed on the left, while communities are displayed on the right. The width of a flow connecting an owner and a community is proportional to the number of outlets in the community belonging to that owner.
\end{flushleft}
\label{comparison_mhash}
\end{figure}

Table \ref{table:clust_perf_eval} shows the results of these indices over the
communities obtained from our similarity graphs. Once again, we can see that
the vocabulary-based similarity does not give a good prediction on news outlets
that belong to the same owner: {\em Vocabulary} gets the lowest score for all
metrics. This poor behavior is shown in a more graphical way in Fig \ref{comparison_vocabulary}. On the other hand, we can see that topic similarities do show
higher degrees of correspondence with owner classes, which is consistent with
previous observations. Keyword-based similarity communities have high homogeneity, while Minhash-based communities show very high completeness, shown in Figs \ref{comparison_topic} and \ref{comparison_mhash}.

We also included in Table \ref{table:clust_perf_eval} the results for indices over the normalized cut clustering obtained for the {\em topic (MinHash-based)} similarity. We follow the same procedure for clusters, i.e., we removed outlets without ownership information and outlets with owners that only have a single outlet. Also, we moved each remaining outlet in cluster ID 0 to its own individual cluster. The completeness (\textbf{Com}) is altered by the bigger number of clusters of size one created from outlets in the cluster with ID 0. The results for this variation (clusters over minhash-based similarity) are the highest for most indices, presenting this technique as a very good predictor for a common-owner relationship.

\FloatBarrier

\section*{Conclusions}
We introduced an analysis of Chilean news media outlets based on the information that each news source chooses to post in Twitter and the similarity clusters of news outlets that arise from this content. The study whether ownership influences the content produced by different news sources is an important area of research to make possible biases explicit.

In general, our results indicate that ownership does play an important role in news content similarity. Big, national-scope media with big audiences tend to group together in their own community; other outlets, generally with a more local scope, group according to a mix of ownership and some geographical features that we inferred (such as {\em Mercurio de Valpara\'iso}).

We studied several similarity metrics as well as different ways in which to identify clusters (or communities) in our data.
The indices that we calculated (e.g., {\bf ARI} and {\bf AMI}) suggest non-random clusters. 

These results seem in agreement with our hypothesis, since similarity based on vocabulary may be attributed to other factors (e.g. geographic zones); on the other hand, our other analyses indicate a correlation between owners and their selection of topics (which can be interpreted as a common editorial policy).

We show that our results are consistent over time. We study two non-overlapping in time datasets, {\em ds15} and {\em ds16}, and show that they both present consistent properties.

A limitation of our methodology is that when outlets are too specialized (e.g. magazines on automobiles, fashion, etc.), even if they belong to the same owner, they do not cluster together. This is due to the nature of the stories that they publish, since by design they do not share any significant part of their content. It is therefore difficult to conclude that there is any influence by owners in the content generation of these specific type of outlets.

Some owners only have one news outlet. As we saw in the \nameref{sec:results} section, our similarity measures based on topics do a good job in pulling apart these special cases from the biggest groups of owners by clustering them together in a different community (our cluster with \textit{id} 0 above). What we are after, however, is the identification of more than one outlet with a single owner such that content in those different media may be affected by a single editorial line.

We have shown that using a language-independent and fast approach we are able to easily discover ``editorial homophily''. This can have several applications, such as help mitigate the  ``filter bubble effect'' in people's news media consumption by recommending more diverse news sources. It can also help identify ``hidden owners'', in the sense that we can identify news sources that behave as if they had a same owner, despite not declaring so publicly. Overall, our findings could help towards promoting a media structure that is less biased towards very few groups.

\pagebreak
\section*{Supporting information}

\begin{table}[ht]
\footnotesize 
\caption{{\bf News outlets for Vocabulary-based communities for the {\em ds16} dataset.}}
\begin{tabular}{|p{0.1\linewidth}|p{0.05\linewidth}|p{0.67\linewidth}|}
\hline
Com. ID &   Size & Outlets  \\
\hline
0 & 79 & diarioelcomunal, sabrosia, elandacollino, primordialfm, radiosolchile, candela\_fm, cappissima, realcondorito ... \\
\hline
1 & 3 & elvicunense, elpaihuanino, elquiglobal \\
\hline
2 & 91 & cooperativa, canal\_13c, nacioncl, t13, bolido\_com, cosmochile, chilebcl, uchileradio, platosycopas, mt\_motore ...\\
\hline
3 & 4 & diariodeaysen, radiolasnieves, patagondomingo, ddivisadero \\
\hline
4 & 43 & austral\_osorno, radiovalparaiso, elrepuertero, soysanantonio, soytome, radio\_festival, ellanquihue ... \\
\hline
5 & 2 & 40chileoficial, fmok \\
\hline
6 & 23 & el\_timeline, antofagastatv, diarioafta, redarica, diariosenred, red\_coquimbo, redantofagasta, soyantofagasta ... \\
\hline
7 & 78 & vallenardigital, radionuble, rsbchile, el\_serenense, elrancaguino, austral\_losrios, laopinon, ultimahoracl ...\\
\hline
8 & 2 & radioeme, lavozdemaipu \\
\hline
9 & 6 & eldia\_cl, diariolabrador, somosmelipilla, portaldemeli, elcomunicadorcl, diarioelcentro \\
\hline
10 & 2 & chilemosaico, proclamacion \\
\hline
11 & 2 & laperladelimari, diarioovallehoy \\
\hline
12 & 2 & prensatuciudad, rengonotas \\
\hline
13 & 2 & mundoacuicola, aquasocial \\
\hline
14 & 2 & radiocorazonfm,radiopudahuel \\
\hline
\end{tabular}
\begin{flushleft} The cluster with an ID 0 corresponds to un-grouped media
outlets.
\end{flushleft}
\label{table:vocab_nps2016} 
\end{table}

\begin{table}[ht]
\footnotesize 
\caption{{\bf Ownership properties for Vocabulary-based clusters for the {\em ds16} dataset.}}
\begin{tabular}{|r|r|p{5cm}|r|r|}
\hline
Com. ID &   Size & Main owners & Owner \% & Unknown owner \%  \\
\hline
\multirow{1}{*}{0} &  \multirow{1}{*}{3} &  el conquistador fm & 33.33 & \multirow{1}{*}{66.67}\\
\hline
\multirow{2}{*}{1} &  \multirow{2}{*}{11} &  el mercurio & 18.18 & \multirow{2}{*}{54.55}\\
\cline{3-4}
 &  & asesorias e inversiones comunidades ciudadanas & 18.18 & \\
\hline
\multirow{8}{*}{2} &  \multirow{8}{*}{9} &  sociedad radiodifusora y periodistica del maule & 11.11 & \multirow{8}{*}{11.11}\\
\cline{3-4}
 &  & radiodifusora paloma & 11.11 & \\
\cline{3-4}
 &  & portales regionales & 11.11 & \\
\cline{3-4}
 &  & grupo diarios en red & 11.11 & \\
\cline{3-4}
 &  & asesorias e inversiones comunidades ciudadanas & 11.11 & \\
\cline{3-4}
 &  & luis verdejo vega & 11.11 & \\
\cline{3-4}
 &  & empresa periodistica curico  & 11.11 & \\
\cline{3-4}
 &  & sociedad radiodifusora cheis & 11.11 & \\
\hline
\multirow{1}{*}{3} &  \multirow{1}{*}{16} &  el mercurio & 100.00 & \multirow{1}{*}{0.00}\\
\hline
\multirow{3}{*}{4} &  \multirow{3}{*}{41} &  copesa & 21.95 & \multirow{3}{*}{0.00}\\
\cline{3-4}
 &  & grupo metro internacional & 12.20 & \\
\cline{3-4}
 &  & editorial televisa chile & 12.20 & \\
\hline
\multirow{1}{*}{5} &  \multirow{1}{*}{12} &  grupo diarios en red & 100.00 & \multirow{1}{*}{0.00}\\
\hline
\multirow{2}{*}{6} &  \multirow{2}{*}{28} &  el mercurio & 14.29 & \multirow{2}{*}{21.43}\\
\cline{3-4}
 &  & asesorias e inversiones comunidades ciudadanas & 25.00 & \\
\hline
\multirow{5}{*}{7} &  \multirow{5}{*}{6} &  medios de consorcio periodistico el epicentro & 16.67 & \multirow{5}{*}{0.00}\\
\cline{3-4}
 &  & el mercurio & 33.33 & \\
\cline{3-4}
 &  & corporacion de television de la pontificia universidad catolica de valparaiso & 16.67 & \\
\cline{3-4}
 &  & comunicaciones pacifico & 16.67 & \\
\cline{3-4}
 &  & asesorias e inversiones comunidades ciudadanas & 16.67 & \\
\hline
\multirow{2}{*}{8} &  \multirow{2}{*}{19} &  el mercurio & 21.05 & \multirow{2}{*}{21.05}\\
\cline{3-4}
 &  & universidad de concepcion & 10.53 & \\
\hline
\multirow{1}{*}{9} &  \multirow{1}{*}{38} &  copesa & 13.16 & \multirow{1}{*}{10.53}\\
\hline
\multirow{2}{*}{10} &  \multirow{2}{*}{4} &  sociedad periodistica e impresora el labrador & 25.00 & \multirow{2}{*}{50.00}\\
\cline{3-4}
 &  & portal de melipilla  & 25.00 & \\
\hline
\multirow{2}{*}{11} &  \multirow{2}{*}{36} &  inversiones canal 13 & 11.11 & \multirow{2}{*}{13.89}\\
\cline{3-4}
 &  & grupo prisa & 13.89 & \\
\hline
12 &  17 &  -- &  -- & 29.41\\
\hline
\multirow{1}{*}{13} &  \multirow{1}{*}{10} &  el mercurio & 40.00 & \multirow{1}{*}{40.00}\\
\hline
\multirow{1}{*}{14} &  \multirow{1}{*}{12} &  red de diarios comunales & 41.67 & \multirow{1}{*}{8.33}\\
\hline
15 &  66 &  -- &  -- & 24.24\\
\hline
\multirow{5}{*}{16} &  \multirow{5}{*}{8} &  el mercurio & 12.50 & \multirow{5}{*}{37.50}\\
\cline{3-4}
 &  & gestion y comunicaciones san lorenzo & 12.50 & \\
\cline{3-4}
 &  & alberto bichara & 12.50 & \\
\cline{3-4}
 &  & enfoque digital & 12.50 & \\
\cline{3-4}
 &  & asesorias e inversiones comunidades ciudadanas & 12.50 & \\
\hline
\multirow{3}{*}{17} &  \multirow{3}{*}{4} &  fundacion democracia y desarrollo & 25.00 & \multirow{3}{*}{25.00}\\
\cline{3-4}
 &  & el democrata & 25.00 & \\
\cline{3-4}
 &  & sociedad periodistica el libero & 25.00 & \\
\hline
\end{tabular}
\begin{flushleft} The cluster with an ID of 15 corresponds to un-grouped media
outlets. Entities owning over 10\% of the outlets in a community are listed next
to it.
\end{flushleft}
\label{table:cvocab_ownership2016} 
\end{table}

\begin{table}[ht]
\small 
\caption{{\bf Ownership properties for Vocabulary-based clusters for the {\em ds15} dataset.}}
\begin{tabular}{|r|r|p{5cm}|r|r|}
\hline
Com. ID &   Size & Main owners & Owner \% & Unknown owner \%  \\
\hline
\multirow{2}{*}{0} &  \multirow{2}{*}{14} &  el mercurio & 57.14 & \multirow{2}{*}{7.14}\\
\cline{3-4}
 &  & estado de chile & 14.29 & \\
\hline
\multirow{2}{*}{1} &  \multirow{2}{*}{4} &  sociedad informativa regional & 25.00 & \multirow{2}{*}{50.00}\\
\cline{3-4}
 &  & asesorias e inversiones comunidades ciudadanas & 25.00 & \\
\hline
\multirow{8}{*}{2} &  \multirow{8}{*}{9} &  el mercurio & 11.11 & \multirow{8}{*}{11.11}\\
\cline{3-4}
 &  & sociedad periodistica de aysen & 11.11 & \\
\cline{3-4}
 &  & sociedad periodistica el ciudadano & 11.11 & \\
\cline{3-4}
 &  & patagonica publicaciones & 11.11 & \\
\cline{3-4}
 &  & la plaza & 11.11 & \\
\cline{3-4}
 &  & ediciones y publicaciones bobby & 11.11 & \\
\cline{3-4}
 &  & sociedad editora y periodistica la verdad & 11.11 & \\
\cline{3-4}
 &  & empresa de publicaciones la prensa austral & 11.11 & \\
\hline
\multirow{2}{*}{3} &  \multirow{2}{*}{2} &  el mercurio & 50.00 & \multirow{2}{*}{0.00}\\
\cline{3-4}
 &  & universidad de concepcion & 50.00 & \\
\hline
\multirow{2}{*}{4} &  \multirow{2}{*}{3} &  el mercurio & 33.33 & \multirow{2}{*}{33.33}\\
\cline{3-4}
 &  & asesorias e inversiones comunidades ciudadanas & 33.33 & \\
\hline
\multirow{2}{*}{5} &  \multirow{2}{*}{2} &  el mercurio & 50.00 & \multirow{2}{*}{0.00}\\
\cline{3-4}
 &  & asesorias e inversiones comunidades ciudadanas & 50.00 & \\
\hline
\multirow{6}{*}{6} &  \multirow{6}{*}{7} &  el mercurio & 14.29 & \multirow{6}{*}{14.29}\\
\cline{3-4}
 &  & grupo mosciatti & 14.29 & \\
\cline{3-4}
 &  & universidad de concepcion & 14.29 & \\
\cline{3-4}
 &  & asesorias e inversiones comunidades ciudadanas & 14.29 & \\
\cline{3-4}
 &  & sociedad nacional de agricultura & 14.29 & \\
\cline{3-4}
 &  & empresa periodistica curico  & 14.29 & \\
\hline
7 &  11 &  -- &  -- & 0.00\\
\hline
\multirow{2}{*}{8} &  \multirow{2}{*}{3} &  el mercurio & 33.33 & \multirow{2}{*}{33.33}\\
\cline{3-4}
 &  & asesorias e inversiones comunidades ciudadanas & 33.33 & \\
\hline
\multirow{2}{*}{9} &  \multirow{2}{*}{3} &  el mercurio & 33.33 & \multirow{2}{*}{33.33}\\
\cline{3-4}
 &  & asesorias e inversiones comunidades ciudadanas & 33.33 & \\
\hline
\multirow{1}{*}{10} &  \multirow{1}{*}{6} &  asesorias e inversiones comunidades ciudadanas & 100.00 & \multirow{1}{*}{0.00}\\
\hline
\multirow{1}{*}{11} &  \multirow{1}{*}{3} &  el mercurio & 100.00 & \multirow{1}{*}{0.00}\\
\hline
\multirow{3}{*}{12} &  \multirow{3}{*}{4} &  el mercurio & 25.00 & \multirow{3}{*}{25.00}\\
\cline{3-4}
 &  & empresa periodistica el observador & 25.00 & \\
\cline{3-4}
 &  & marcelo jara olivares & 25.00 & \\
\hline
\multirow{6}{*}{13} &  \multirow{6}{*}{8} &  sociedad periodistica e impresora el labrador & 12.50 & \multirow{6}{*}{12.50}\\
\cline{3-4}
 &  & copesa & 12.50 & \\
\cline{3-4}
 &  & asesorias e inversiones comunidades ciudadanas & 12.50 & \\
\cline{3-4}
 &  & carabineros de chile & 12.50 & \\
\cline{3-4}
 &  & editora el centro empresa periodistica & 12.50 & \\
\cline{3-4}
 &  & antonio puga & 25.00 & \\
\hline
\end{tabular}
\begin{flushleft} The cluster with an ID of 7 corresponds to un-grouped media
outlets. Entities owning over 10\% of the outlets in a community are listed next
to it.
\end{flushleft}
\label{table:cvocab_ownership2015} 
\end{table}

\begin{table}[ht]
\footnotesize 
\caption{{\bf News outlets for Topic keyword-based communities for the {\em ds16} dataset.}}
\begin{tabular}{|p{0.1\linewidth}|p{0.05\linewidth}|p{0.68\linewidth}|}
\hline
Com. ID &   Size & Outlets  \\
\hline
0 & 208 & elvicunense, canal\_13c, diariodeaysen, diarioelcomunal, sabrosia, bolido\_com, 40chileoficial ... \\
\hline
1 & 59 & cooperativa, nacioncl, t13, chilebcl, radionuble, eldia\_cl, gamba\_cl, emol, eldesconcierto, lacuarta ...\\
\hline
2 & 14 & austral\_osorno, austral\_losrios, ellanquihue, estrellaconce, clave9cl, diarioatacama, estrelladearica ... \\
\hline
3 & 4 & radiovalparaiso, radio\_festival, elepicentro, ucvradio \\
\hline
4 & 16 & uchileradio, elrepuertero, laopinon, el\_amaule, elvacanudo, elnaveghable, elmagallanews, elobservatodo ... \\
\hline
5 & 25 & soysanantonio, soytome, soyantofagasta, laestrellavalpo, soytemuco, laestrellaiqq, soycopiapo ... \\
\hline
6 & 13 & redarica, diariosenred, red\_coquimbo, redantofagasta, redaraucania, redmaule, redbiobio, redlosrios ... \\
\hline
7 & 2 & putaendoinforma, aconcaguanews \\
\hline
\end{tabular}
\begin{flushleft} The cluster with an ID 0 corresponds to un-grouped media
outlets.
\end{flushleft}
\label{table:topic_nps2016} 
\end{table}

\begin{table}[ht]
\footnotesize 
\caption{{\bf Ownership properties for Topic keyword-based clusters for the {\em ds16} dataset.}}
\begin{tabular}{|r|r|p{5cm}|r|r|}
\hline
Com. ID &   Size & Main owners & Owner \% & Unknown owner \%  \\
\hline
\multirow{1}{*}{0} &  \multirow{1}{*}{13} &  grupo diarios en red & 100.00 & \multirow{1}{*}{0.00}\\
\hline
1 &  158 &  -- &  -- & 27.85\\
\hline
\multirow{1}{*}{2} &  \multirow{1}{*}{1} &  el mercurio & 100.00 & \multirow{1}{*}{0.00}\\
\hline
\multirow{1}{*}{3} &  \multirow{1}{*}{15} &  asesorias e inversiones comunidades ciudadanas & 100.00 & \multirow{1}{*}{0.00}\\
\hline
\multirow{1}{*}{4} &  \multirow{1}{*}{12} &  el mercurio & 100.00 & \multirow{1}{*}{0.00}\\
\hline
5 &  13 &  -- &  -- & 0.00\\
\hline
\multirow{4}{*}{6} &  \multirow{4}{*}{5} &  medios de consorcio periodistico el epicentro & 20.00 & \multirow{4}{*}{0.00}\\
\cline{3-4}
 &  & el mercurio & 40.00 & \\
\cline{3-4}
 &  & corporacion de television de la pontificia universidad catolica de valparaiso & 20.00 & \\
\cline{3-4}
 &  & comunicaciones pacifico & 20.00 & \\
\hline
\multirow{2}{*}{7} &  \multirow{2}{*}{30} &  copesa & 13.33 & \multirow{2}{*}{3.33}\\
\cline{3-4}
 &  & grupo prisa & 20.00 & \\
\hline
\multirow{1}{*}{8} &  \multirow{1}{*}{3} &  el mercurio & 100.00 & \multirow{1}{*}{0.00}\\
\hline
\multirow{4}{*}{9} &  \multirow{4}{*}{5} &  sociedad de comunicaciones el trabajo & 20.00 & \multirow{4}{*}{20.00}\\
\cline{3-4}
 &  & marcelo jara olivares & 20.00 & \\
\cline{3-4}
 &  & sociedad radio aconcagua  & 20.00 & \\
\cline{3-4}
 &  & patricio gallardo montenegro & 20.00 & \\
\hline
\multirow{1}{*}{10} &  \multirow{1}{*}{23} &  copesa & 13.04 & \multirow{1}{*}{17.39}\\
\hline
11 &  34 &  -- &  -- & 20.59\\
\hline
\multirow{1}{*}{12} &  \multirow{1}{*}{16} &  el mercurio & 100.00 & \multirow{1}{*}{0.00}\\
\hline
\end{tabular}
\begin{flushleft} The cluster with an ID of 1 corresponds to un-grouped media
outlets. Entities owning over 10\% of the outlets in a community are listed next
to it.
\end{flushleft}
\label{table:ctopic_ownership2016} 
\end{table}

\begin{table}[ht]
\footnotesize 
\caption{{\bf Ownership properties for Topic keyword-based clusters for the {\em ds15} dataset.}}
\begin{tabular}{|r|r|p{5cm}|r|r|}
\hline
Com. ID &   Size & Main owners & Owner \% & Unknown owner \%  \\
\hline
\multirow{9}{*}{0} &  \multirow{9}{*}{9} &  el mercurio & 11.11 & \multirow{9}{*}{0.00}\\
\cline{3-4}
 &  & estado de chile & 11.11 & \\
\cline{3-4}
 &  & grupo mosciatti & 11.11 & \\
\cline{3-4}
 &  & copesa & 11.11 & \\
\cline{3-4}
 &  & terra networks chile & 11.11 & \\
\cline{3-4}
 &  & fundacion para las comunicaciones sociales & 11.11 & \\
\cline{3-4}
 &  & comunicaciones lanet & 11.11 & \\
\cline{3-4}
 &  & salvador schwartzmann & 11.11 & \\
\cline{3-4}
 &  & inversiones canal 13 & 11.11 & \\
\hline
1 &  28 &  -- &  -- & 32.14\\
\hline
\multirow{1}{*}{2} &  \multirow{1}{*}{5} &  el mercurio & 100.00 & \multirow{1}{*}{0.00}\\
\hline
\multirow{1}{*}{3} &  \multirow{1}{*}{7} &  el mercurio & 100.00 & \multirow{1}{*}{0.00}\\
\hline
\multirow{1}{*}{4} &  \multirow{1}{*}{12} &  asesorias e inversiones comunidades ciudadanas & 100.00 & \multirow{1}{*}{0.00}\\
\hline
\multirow{1}{*}{5} &  \multirow{1}{*}{3} &  el mercurio & 100.00 & \multirow{1}{*}{0.00}\\
\hline
\multirow{1}{*}{6} &  \multirow{1}{*}{2} &  asesorias e inversiones comunidades ciudadanas & 100.00 & \multirow{1}{*}{0.00}\\
\hline
\multirow{6}{*}{7} &  \multirow{6}{*}{7} &  el mercurio & 14.29 & \multirow{6}{*}{14.29}\\
\cline{3-4}
 &  & cnn chile & 14.29 & \\
\cline{3-4}
 &  & sociedad periodistica el ciudadano & 14.29 & \\
\cline{3-4}
 &  & ediciones y publicaciones bobby & 14.29 & \\
\cline{3-4}
 &  & universidad de concepcion & 14.29 & \\
\cline{3-4}
 &  & la plaza & 14.29 & \\
\hline
\multirow{1}{*}{8} &  \multirow{1}{*}{2} &  el mercurio & 100.00 & \multirow{1}{*}{0.00}\\
\hline
\multirow{4}{*}{9} &  \multirow{4}{*}{4} &  copesa & 25.00 & \multirow{4}{*}{0.00}\\
\cline{3-4}
 &  & grupo bethia & 25.00 & \\
\cline{3-4}
 &  & grupo metro internacional & 25.00 & \\
\cline{3-4}
 &  & grupo prisa & 25.00 & \\
\hline
\end{tabular}
\begin{flushleft} The cluster with an ID of 1 corresponds to un-grouped media
outlets. Entities owning over 10\% of the outlets in a community are listed next
to it.
\end{flushleft}
\label{table:ctopic_ownership2015} 
\end{table}

\begin{table}[ht]
\footnotesize 
\caption{{\bf News outlets for Topic minhash-based communities for the {\em ds16} dataset.}}
\begin{tabular}{|p{0.1\linewidth}|p{0.05\linewidth}|p{0.68\linewidth}|}
\hline
Com. ID &   Size & Outlets  \\
\hline
0 & 10 & ahoranoticiasan, carabdechile, chilebcl, diarioenaccion, futurafmoficial, hoyxhoycl, la\_segunda ... \\
\hline
1 & 109 & 24horastvn, 40chileoficial, adnradiochile, antofacity\_com, armoniaonline, biobio, biobiodeportivo ... \\
\hline
2 & 43 & 33temuco, austral\_losrios, austral\_osorno, australtemuco, cronicachillan, diarioatacama, diarioelhuemul ... \\
\hline
3 & 130 & acciondeongs, aconcaguanews, aconcaguaradio, agriculturafm, alfaomegacurico, americaeconomia, antofagastatv ... \\
\hline
4 & 2 & ancoafm, canal5linares \\
\hline
5 & 2 & aquasocial, mundoacuicola \\
\hline
6 & 9 & betazeta, bolido\_com, chw\_net, fayerwayer, ferplei, niubie\_com, sabrosia, veoverde, wayerless \\
\hline
7 & 36 & carta\_abierta, cnnchile, concordia\_arica, diario\_eha, diariolabrador, el\_amaule, el\_provincial ... \\
\hline
8 & 6 & chiloealdia, elinsular1, estrellachiloe, lacuarta, queilencl, radioquellon \\
\hline
9 & 14 & diariosenred, elliberocl, red\_coquimbo, red\_ohiggins, redantofagasta, redaraucania, redarica, redatacama ... \\
\hline
10 & 2 & lavozdemaipu, radioeme \\
\hline
11 & 2 & primordialfm, ultimahoracl \\
\hline
\end{tabular}
\begin{flushleft} The cluster with an ID 0 corresponds to un-grouped media
outlets.
\end{flushleft}
\label{table:minhash_nps2016} 
\end{table}

\begin{table}[ht]
\footnotesize 
\caption{{\bf Ownership properties for Topic minhash-based clusters for the {\em ds15} dataset.}}
\begin{tabular}{|r|r|p{5cm}|r|r|}
\hline
Com. ID &   Size & Main owners & Owner \% & Unknown owner \%  \\
\hline
0 &  31 &  -- &  -- & 22.58\\
\hline
\multirow{1}{*}{1} &  \multirow{1}{*}{14} &  el mercurio & 100.00 & \multirow{1}{*}{0.00}\\
\hline
\multirow{1}{*}{2} &  \multirow{1}{*}{11} &  asesorias e inversiones comunidades ciudadanas & 100.00 & \multirow{1}{*}{0.00}\\
\hline
\multirow{1}{*}{3} &  \multirow{1}{*}{1} &  el mercurio & 100.00 & \multirow{1}{*}{0.00}\\
\hline
\multirow{1}{*}{4} &  \multirow{1}{*}{1} &  grupo mosciatti & 100.00 & \multirow{1}{*}{0.00}\\
\hline
\multirow{1}{*}{5} &  \multirow{1}{*}{2} &  copesa & 50.00 & \multirow{1}{*}{50.00}\\
\hline
\multirow{3}{*}{6} &  \multirow{3}{*}{3} &  alberto bichara & 33.33 & \multirow{3}{*}{0.00}\\
\cline{3-4}
 &  & universidad de concepcion & 33.33 & \\
\cline{3-4}
 &  & empresa de publicaciones la prensa austral & 33.33 & \\
\hline
\multirow{3}{*}{7} &  \multirow{3}{*}{3} &  sociedad periodistica e impresora el labrador & 33.33 & \multirow{3}{*}{0.00}\\
\cline{3-4}
 &  & grupo metro internacional & 33.33 & \\
\cline{3-4}
 &  & grupo bethia & 33.33 & \\
\hline
\multirow{3}{*}{8} &  \multirow{3}{*}{4} &  marcelo jara olivares & 25.00 & \multirow{3}{*}{0.00}\\
\cline{3-4}
 &  & terra networks chile & 25.00 & \\
\cline{3-4}
 &  & asesorias e inversiones comunidades ciudadanas & 50.00 & \\
\hline
\multirow{1}{*}{9} &  \multirow{1}{*}{2} &  antonio puga & 100.00 & \multirow{1}{*}{0.00}\\
\hline
\multirow{1}{*}{10} &  \multirow{1}{*}{2} &  estado de chile & 100.00 & \multirow{1}{*}{0.00}\\
\hline
\multirow{2}{*}{11} &  \multirow{2}{*}{4} &  sociedad informativa regional & 25.00 & \multirow{2}{*}{50.00}\\
\cline{3-4}
 &  & asesorias e inversiones comunidades ciudadanas & 25.00 & \\
\hline
\multirow{1}{*}{12} &  \multirow{1}{*}{1} &  fundacion para las comunicaciones sociales & 100.00 & \multirow{1}{*}{0.00}\\
\hline
\multirow{1}{*}{13} &  \multirow{1}{*}{2} &  el mercurio & 100.00 & \multirow{1}{*}{0.00}\\
\hline
\end{tabular}
\begin{flushleft} The cluster with an ID of 0 corresponds to un-grouped media
outlets. Entities owning over 10\% of the outlets in a community are listed next
to it.
\end{flushleft}
\label{table:cminhash_ownership2015} 
\end{table}

\FloatBarrier

\section*{Acknowledgments}

EE was supported by the doctoral scholarships of CONICYT No. 63130228. LF would like to thank Movistar - Telef\'onica Chile and the Chilean government initiative CORFO 13CEE2-21592 (2013-21592-1-INNOVA\_PRODUCCION2013-21592-1) for financial support. Funded, in part, by Conicyt's Proyecto de Informacion Cientifica, Pluralismo en el Sistema Informativo Nacional, PLU140001. We thank Nicolás del Valle, and his database of news media through Proyecto Pluralismo PLU1300008, and Miguel Paz for his Mapa de Medios by Poderopedia. 
\section*{Author Contributions}
All authors were involved in all stages of this research.

%
%
%


\begin{thebibliography}{10}

\bibitem{10.1371/journal.pone.0014243}
Ilias Flaounas, Marco Turchi, Omar Ali, Nick Fyson, Tijl De~Bie, Nick Mosdell,
  Justin Lewis, and Nello Cristianini.
\newblock The structure of the eu mediasphere.
\newblock {\em PLoS ONE}, 5(12):e14243, 12 2010.

\bibitem{pariser2011filter}
Eli Pariser.
\newblock The Filter Bubble: What the Internet Is Hiding from You.
\newblock The Penguin Group, 2011. ISBN 1594203008.

\bibitem{bagdikian2004new}
Ben H. Bagdikian.
\newblock The Media Monopoly, 6th Edition.
\newblock Beacon Press, 2000. ISBN 9780807061794.

\bibitem{bakshy2015exposure}
Eytan Bakshy, Solomon Messing, and Lada A Adamic
\newblock Exposure to ideologically diverse news and opinion on Facebook.
\newblock {\em Science}, 348(6239), pages 1130--1132, 2015

\bibitem{10.1371/journal.pone.0061981}
Delia Mocanu, Andrea Baronchelli, Nicola Perra, Bruno Gonçalves, Qian Zhang,
  and Alessandro Vespignani.
\newblock The twitter of babel: Mapping world languages through microblogging
  platforms.
\newblock {\em PLoS ONE}, 8(4):e61981, 04 2013.

\bibitem{manber1994finding}
Udi Manber et~al.
\newblock Finding similar files in a large file system.
\newblock In {\em Usenix Winter}, volume~94, pages 1--10, 1994.

\bibitem{broder1997resemblance}
Andrei~Z Broder.
\newblock On the resemblance and containment of documents.
\newblock In {\em Compression and Complexity of Sequences 1997. Proceedings},
  pages 21--29. IEEE, 1997.
  
\bibitem{broder2000identifying}
Andrei~Z Broder.
\newblock Identifying and filtering near-duplicate documents.
\newblock In {\em Combinatorial pattern matching}, pages 1--10. Springer, 2000.

\bibitem{manku2007detecting}
Gurmeet~Singh Manku, Arvind Jain, and Anish Das~Sarma.
\newblock Detecting near-duplicates for web crawling.
\newblock In {\em Proceedings of the 16th international conference on World
  Wide Web}, pages 141--150. ACM, 2007.

\bibitem{rajaraman2012mining}
Anand Rajaraman, Jeffrey~D Ullman, Jeffrey~David Ullman, and Jeffrey~David
  Ullman.
\newblock {\em Mining of massive datasets}, volume~77.
\newblock Cambridge University Press Cambridge, 2012.

\bibitem{broder1998min}
Andrei~Z Broder, Moses Charikar, Alan~M Frieze, and Michael Mitzenmacher.
\newblock Min-wise independent permutations.
\newblock In {\em Proceedings of the thirtieth annual ACM symposium on Theory
  of computing}, pages 327--336. ACM, 1998.
  
\bibitem{buehrer2008scalable}
Gregory Buehrer and Kumar Chellapilla.
\newblock A scalable pattern mining approach to web graph compression with
  communities.
\newblock In {\em Proceedings of the 2008 International Conference on Web
  Search and Data Mining}, pages 95--106. ACM, 2008.
  
\bibitem{hernandez2014compressed}
Cecilia Hern{\'a}ndez and Gonzalo Navarro.
\newblock Compressed representations for web and social graphs.
\newblock {\em Knowledge and information systems}, 40(2):279--313, 2014.

\bibitem{chierichetti2009compressing}
Flavio Chierichetti, Ravi Kumar, Silvio Lattanzi, Michael Mitzenmacher,
  Alessandro Panconesi, and Prabhakar Raghavan.
\newblock On compressing social networks.
\newblock In {\em Proceedings of the 15th ACM SIGKDD international conference
  on Knowledge discovery and data mining}, pages 219--228. ACM, 2009.
  
\bibitem{urvoy2008tracking}
Tanguy Urvoy, Emmanuel Chauveau, Pascal Filoche, and Thomas Lavergne.
\newblock Tracking web spam with html style similarities.
\newblock {\em ACM Transactions on the Web (TWEB)}, 2(1):3, 2008.

\bibitem{berlin2015assembling}
Konstantin Berlin, Sergey Koren, Chen-Shan Chin, James~P Drake, Jane~M
  Landolin, and Adam~M Phillippy.
\newblock Assembling large genomes with single-molecule sequencing and
  locality-sensitive hashing.
\newblock {\em Nature biotechnology}, 2015.

\bibitem{Saez-Trumper:2013:SMN:2505515.2505623}
Diego Saez-Trumper, Carlos Castillo, and Mounia Lalmas.
\newblock Social media news communities: Gatekeeping, coverage, and statement
  bias.
\newblock In {\em Proceedings of the 22Nd ACM International Conference on
  Information \& Knowledge Management}, CIKM '13, pages 1679--1684, New York,
  NY, USA, 2013. ACM.
  
\bibitem{citeulike:9609587}
Jisun An, Meeyoung Cha, Krishna Gummadi, and Jon Crowcroft.
\newblock Media Landscape in {Twitter}: A World of New Conventions and Political Diversity.
\newblock In {\em Proceedings of the Fifth International Conference on Weblogs and Social Media},
Menlo Park, CA, USA, 2011. AAAI.
  
\bibitem{Shi:2000:NCI:351581.351611}
Jianbo Shi, and Jitendra Malik.
\newblock Normalized Cuts and Image Segmentation.
\newblock {\em IEEE Trans. Pattern Anal. Mach. Intell.}, 22(8):888--905, 2000.

\bibitem{Thorndike1953}
Robert L. Thorndike.
\newblock Who belongs in the family?.
\newblock {\em Psychometrika}, 18(4):267--276, 1953. 

\bibitem{Park:2012:CFM:2209310.2209311}
Souneil Park, Seungwoo Kang, Sangyoung Chung, and Junehwa Song.
\newblock A computational framework for media bias mitigation.
\newblock {\em ACM Trans. Interact. Intell. Syst.}, 2(2):8:1--8:32, June 2012.


\bibitem{watts2002factor}
Watts RJ, Porter AL, Zhu D.
\newblock Factor Analysis Optimization: Applied on Natural Language Knowledge
  Discovery.
\newblock In: In: Committee on Data for Science and Technology 2002: Frontiers
  of Scientific and Technical Data: Proceedings of the 18 th International
  Conference CODATA 2002; 2002. Available from:
  \url{{http://www.dtic.mil/cgi-bin/GetTRDoc?Location=U2&doc=GetTRDoc.pdf&AD=ADA484817}}.

\bibitem{almeida2011there}
Almeida H, Neto DOG, Jr WM, Zaki MJ.
\newblock Is There a Best Quality Metric for Graph Clusters?
\newblock In: Machine Learning and Knowledge Discovery in Databases - European
  Conference, {ECML} {PKDD} 2011, Athens, Greece, September 5-9, 2011.
  Proceedings, Part {I}; 2011. p. 44--59.
\newblock Available from: \url{http://dx.doi.org/10.1007/978-3-642-23780-5_13}.

\bibitem{greedymod}
{Clauset} A, {Newman} MEJ, {Moore} C.
\newblock Finding community structure in very large networks.
\newblock Physical Review E. 2004 Dec;70(6).
\newblock Available from: \url{http://dx.doi.org/10.1103/PhysRevE.70.066111}.

\bibitem{twitterglossary}
{Twitter, Inc }. The {{Twitter}} glossary; 2016.
\newblock Accesed: 2016-04-02.
\newblock \url{https://support.twitter.com/articles/166337}.

\bibitem{Castro2008}
Castro C.
\newblock {Industrias de Contenidos en Latinoam\'{e}rica}.
\newblock Meta. 2008;Available from:
  \url{http://www.razonypalabra.org.mx/libros/libros/Gdt_eLAC_meta_13.pdf}.

\bibitem{manning2008scoring}
Manning CD, Raghavan P, Sch{\"u}tze H.
\newblock 6 - Scoring, term weighting and the vector space model.
\newblock In: Introduction to Information Retrieval. Cambridge University
  Press; 2008. .

\bibitem{mquezada}
Kalyanam J, Quezada M, Poblete B, Lanckriet GRG.
\newblock Early prediction and characterization of high-impact world events
  using social media.
\newblock CoRR. 2015;abs/1511.01830.
\newblock Available from: \url{http://arxiv.org/abs/1511.01830}.

\bibitem{kurt}
DeCarlo LT.
\newblock On the meaning and use of kurtosis.
\newblock Psychological methods. 1997;2(3):292.
\newblock Available from: \url{http://dx.doi.org/10.1037/1082-989X.2.3.292}.

\bibitem{graphvis}
Freeman LC.
\newblock Visualizing social networks.
\newblock Journal of social structure. 2000;1(1):4.
\newblock Available from:
  \url{http://www.cmu.edu/joss/content/articles/volume1/Freeman.html}.


\bibitem{jaz}
Maldonado J, Peña-Araya V, Poblete B.
\newblock Spatio and Temporal Characterization of Chilean News Events in Social
  Media.
\newblock In: SIGIR 2015 Workshop on Temporal, Social and Spatially-aware
  Information Access (TAIA '15); 2015. 

\bibitem{singhal2001modern}
Singhal A.
\newblock Modern Information Retrieval: {A} Brief Overview.
\newblock {IEEE} Data Eng Bull. 2001;24(4):35--43.
\newblock Available from: \url{http://sites.computer.org/debull/A01DEC-CD.pdf}.


\bibitem{newman2004}
Newman, M. E. J.
\newblock Fast algorithm for detecting community structure in networks.
\newblock {APS} Physics Review E. 2004;69(6): 066133.
\newblock Available from: \url{http://arxiv.org/pdf/cond-mat/0309508.pdf}.

\bibitem{Lu:2015:BLS:2806416.2806573}
Haokai Lu, James Caverlee, and Wei Niu
\newblock BiasWatch: A Lightweight System for Discovering and Tracking Topic-Sensitive Opinion Bias in Social Media.
\newblock In Proceedings of the 24th ACM International on Conference on Information and Knowledge Management (CIKM '15).
\newblock ACM, New York, NY, USA, 213-222. DOI=http://dx.doi.org/10.1145/2806416.2806573

\bibitem{chomsky1998common}
Chomsky, N. and Barsamian, D. and Naiman, A.
\newblock The common good.
\newblock Real story series.
\newblock Odonian Press; 1998.
\newblock Available from: \url{https://books.google.de/books?id=2ZK6AAAAIAAJ}.

\bibitem{chomsky1988consent}
Herman, E. S. and Chomsky, N.
\newblock Manufacturing consent : the political economy of the mass media.
\newblock New York: Pantheon Books, 1988.

\bibitem{messner2011shovel}
Messner, Marcus, Maureen Linke, and Asriel Eford. 
\newblock Shoveling tweets: An analysis of the microblogging engagement of traditional news organizations.
\newblock International Symposium on Online Journalism in Austin, 2011.
\newblock Available from: \url{http://www.academia.edu/2860620/Shoveling\_tweets\_An\_analysis\_of\_the \_microblogging\_engagement\_of\_traditional\_news\_organizations}


\bibitem{Winseck2008}
Winseck, D. 
\newblock The State of Media Ownership and Media Markets: Competition or Concentration and Why Should We Care?. 
\newblock Sociology Compass 2008, 2: 34–47. doi:10.1111/j.1751-9020.2007.00061.x

\bibitem{twitter}
Company | About. Twitter. Twitter; Available: https://about.twitter.com/company

\bibitem{wiki.chilean.newspapers}
Medios de comunicación en Chile. Wikipedia. Wikimedia Foundation; 2017. Available: https://es.wikipedia.org/wiki/Medios\_de\_comunicaci\%C3\%B3n\_en\_Chile

\bibitem{poderopedia}
Mapa de Medios | Poderopedia.[cited 12 May 2017] Poderomedia Foundation; 2017. Available from: http://apps.poderopedia.org/mapademedios/index/
\end{thebibliography}
\end{document}